\documentstyle[hip-artc,epsf]{article}
\volnumber{5}  \edyear{1997}  \frompage{255} \topage{269}                
\recrevdate{8 April 1997}
\title{QCD Spectra and Random Matrix Models}
\authors{
{\twerm 
  G\'abor Papp$^{1,2}$
}\\[2.812mm]
{\normalsize
\hspace*{-8pt}$^1$ Gesellschaft f\"ur Schwerionenforschung,\\ 
 D-64291 Darmstadt, Germany\\[0.2ex]
\hspace*{-8pt}$^2$ Institute for Theoretical Physics, E\"{o}tv\"{o}s
 University, \\
 H-1088 Budapest, Hungary\\[0.2ex]
}}

\abstract{%
We summarize some recent results on the application of macroscopic
spectral properties of random matrix models (RMM) to the QCD spectra. A
comparison to existing lattice simulation is presented both for staggered 
and Wilson fermions for high but finite temperature. We consider two
type of mixing between the four lowest Matsubara modes, corresponding to
third and fifth order algebraic equation for the pertinent resolvent, 
respectively.}

\newcommand{\be}{\begin{eqnarray}}
\newcommand{\ee}{\end{eqnarray}}

\begin{document}

\maketitle

\section{Introduction}

Dedicated lattice simulations on the aspects of the QCD phase transition
have triggered a number of theoretical investigations aimed at understanding
some of the nonperturbative aspects of QCD whether in vacuum or matter 
\cite{KANAYA}. In particular, it was shown recently that the bulk character
of the QCD Dirac spectra could be understood using simple random matrix models
\cite{WILSON,RANDOM,BLUE}, and that the spectral-level correlations are
consistent with the general lore of random matrix theory
\cite{HALAS,MICRO}.

At zero temperature but for massive quarks, QCD simulations with Wilson
quarks show that the Dirac spectrum exhibits a structural change with
increasing quark  
mass, a feature that is reflected in the random matrix version in terms of 
heavy quarks becoming localized over their Compton wavelength \cite{RANDOM}. 
A similar behavior is also observed at finite temperature when the
quark modes are restricted to the lowest Matsubara modes 
$\omega_0 = |\omega_{-1} | = \pi T$ \cite{WILSON,JACKSON}. 

At high temperature, the static screening lengths in QCD are dominated by the 
lowest Matsubara modes \cite{ZAHEDHANS}. This behavior is confirmed by
lattice  
simulations for temperatures that are remarkably close to the critical 
temperature \cite{DETAR}. At lower temperatures, the role of higher 
Matsubara modes become more relevant. One purpose of this paper, is to 
investigate the effects of few Matsubara modes (four) in random matrix models. 
In Section 2, we use the one-flavor version of the NJL Lagrangian in Euclidean 
space to discuss the effects of mixing between various Matsubara modes in 
random matrix models \cite{USNJL}. In Section 3, we specialize to the case of 
Wilson fermions and derive Pastur's equation \cite{PASTUR}
for the pertinent random matrix model for the two cases where the 
Matsubara modes are either mixed (model I), or decoupled (model II). In 
Section 4, the Dirac spectral distribution for model I is shown to follow 
from a fifth order algebraic equation (Quinto class), while the evolution 
of the end-points follow from a quartic equation. In Section 5, the
Dirac spectrum 
associated to model II is discussed using a superposition of a solution 
to a cubic equation (Cardano class). Some further suggestions and conclusions
are made in Section 6. In Section 7 we give a short overview of the 
applicability of such models to chiral fermions, and particularly we
reproduce the valence quark condensate calculated on lattice~\cite{CHRIST}.
Finally, in Section 8 we summarize.

\section{Mapping of the NJL Model into RMM}

The relevance of the Matsubara modes in random matrix models 
along with their mixing is best seen using one flavor NJL Lagrangian in
Euclidean space, with anti-periodic boundary conditions. Specifically
\be
  {\cal L}_4 =
	{\psi}^{\dagger}( i\gamma\cdot \partial  + im  ) 
	  \psi
  + \frac {g^2}{2} \bigg( (\psi^{\dagger}\psi )^2 + 
	(\psi^{\dagger} i\gamma_5 \psi)^2 \bigg) \,,
\label{toymodel}
\ee
where $g$ is a fixed coupling constant. Using the standard bosonization 
prescription
\be 
P \leftrightarrow -2ig^2\psi_L^{\dagger}\psi_L \quad , \quad
P^{\dagger} \leftrightarrow -2ig^2\psi_R^{\dagger}\psi_R \,,
\label{bozo}
\ee
where $P, P^{\dagger}$ stand for  independent 
auxiliary fields, we may rewrite (\ref{toymodel}) in the form
\be
  {\cal L}_4 &=&+ {\psi}_R^{\dagger}( i\gamma\cdot \partial)
	\psi_L + {\psi}_L^{\dagger}( i\gamma\cdot \partial)
	\psi_R \nonumber\\
  &+& \psi_R^{\dagger}\, i ( P + m) \, \psi_R + \psi^{\dagger}_L i 
	( P^{\dagger} + m) 
	\psi_L  + \frac {1}{2g^2} PP^{\dagger}
\label{PT2}
\ee
in the chiral basis $\psi = (\psi_R , \psi_L )$. 
Note that the Minkowski fields follow from the Euclidean fields through 
$(i\psi^{\dagger} , \psi )\rightarrow (\overline{\psi} , \psi )$. 
Equation (\ref{PT2}) is defined on 
the strip $\beta \times V_3$ in Euclidean space, 
with 
$P(\tau + \beta , \vec x ) = P (\tau, \vec x )$, and $\psi (\tau + \beta , 
\vec x ) = -\psi (\tau, \vec x )$, and a  3-momentum cut-off 
$\Lambda$. 
We simplify further the model by putting the left and right quark fields on 
a discrete grid spanned by the three-space points ${\bf {x}}=1,2,\dots, N$ 
and introducing the rescaled fields $q_n^{\bf x}=\sqrt{V_3}\psi_n^{\bf
x}$ with  
\be
\psi (\tau , \vec{x} ) = \sum_{n=-\infty}^{+\infty} e^{-i\omega_n \tau} 
\,\psi_n^{\bf x} \,,
\label{matsu}
\ee
where $\omega_n= (2n+1) \pi T$ are the Matsubara frequencies ($T=1/\beta $).
In what follows we choose $P$ $\tau-$independent. This is equivalent to 
restricting the four-quark interaction
\be
  \int_0^{\beta}\!d\tau\ \left( (\psi^{\dagger}\psi )^2  + 
	(\psi^{\dagger}i\gamma_5 \psi)^2\right) = 
 4\! \sum_{n,m,k,l}\!\delta_{n+k,m+l} \ 
	\psi^{\dagger}_{Rn} \psi_{Rm} \psi_{Lk}^{\dagger} \psi_{Ll}
\label{X3}
\ee
to only the terms with $n=m$ and $k=l$. Other choices are possible.

In the case where  the auxiliary fields $P$ are $\tau$-independent, 
we can bosonize the present model in a different but complementary way
\cite{USNJL}. After integrating (\ref{PT2}) over $P$ and $P^{\dagger}$, 
we bosonize the chirality-flipping pairs using
\be
{\bf R}_{n,m}^{{\bf x,y}}= q_{Rm}^{\bf x} q^{\dagger {\bf y}}_{Ln}
\label{randomstuff}
\ee
which is a doubly banded, complex matrix ${\bf R}$ with dimensions
$(N\times N)\otimes (\infty \times \infty)$. The upper indices
refer to three-space ${\bf x,y}$ and the lower indices to
frequency space $n,m$. In terms of (\ref{randomstuff}), the NJL Lagrangian 
(\ref{toymodel}) becomes
\be
  {\cal L}_0 =
	q^{\dagger} ( {\bf \Omega} \gamma_4 + im ) q
  +N \Sigma {\rm Tr}_{{\bf x},n} ({\bf R}{\bf R}^{\dagger})
	+q_{R}^{\dagger } {\bf R} q_{L} + q_{L}^{\dagger} 
	{\bf R}^{\dagger}q_{R} \,,
\label{X4}
\ee
where the trace in (\ref{X4}) is over ${\bf x}$ and $n$. 
The partition function 
associated to (\ref{X4}) is simply
\be
Z[T, \mu ] =\int\!d{\bf R}\  
e^{-N\Sigma {\rm Tr}_{{\bf x},n} ({\bf R}{\bf R}^{\dagger} )}\,\,
{\rm det}_{2,{\bf x},n} {\bf Q}
\label{X5}
\ee
with the Dirac operator in a random background,
\be
{\bf Q}_S = \left(\matrix{ im &  {\bf \Omega} \cr
 {\bf \Omega}  & im \cr}\right) + \left(\matrix{ 0 & {\bf R}\cr
{\bf R}^{\dagger} & 0 \cr}\right) \,,
\label{XDIRAC}
\ee
where ${\bf \Omega}=\omega_n{\bf 1}_n \otimes {\bf 1}_{\bf x}$.
The determinant in (\ref{X5}) is over chirality (2), space (${\bf x}$),
and frequency space ($n$). This is  an example of a chiral random matrix 
model \cite{CHIRALRANDOM}. The structure of (\ref{XDIRAC}) generalizes to 
several flavors as discussed in \cite{WILSON}. 
The spectral distribution associated to (\ref{XDIRAC}) exhibits manifest 
chirality, and is suitable for describing
Kogut-Susskind fermions on the lattice, as we will discuss below.

\section{Massive Quark Spectra}

The case of Wilson fermions is in a sense different. Indeed, on the lattice, 
Wilson fermions are not manifestly chirally symmetric due to the presence of 
the r-terms \cite{LATTICEWILSON}. To compare with lattice Wilson spectra, 
we consider instead the hermitean operator
${\bf Q} = \gamma_5 (D\!\!\!\!/ +  m )$, where
${\bf Q}$ is the Dirac $Hamiltonian$ in five-dimensions and a single flavor. 
Here,
the Dirac matrix $\gamma_5$ plays the role of the $\beta$-Dirac matrix
in (3+1) dimensions. A number of simulations, using the analog of ${\bf Q}$ 
with Wilson fermions on the lattice have been carried out recently by 
Kalkreuter\cite{KALKREUTER} for two-color and two-flavor QCD. His results
are in qualitative agreement with random matrix 
theory\cite{RANDOM,BLUE,HALASZJAC}. In~\cite{BLUE} the low energy QCD
spectra was calculated taking into account the lowest Matsubara pair
only. The lattice calculation~\cite{KALKREUTER} however shows the
appearance of at least one more scale. One of the possible sources of
another scale is the presence of the next Matsubara pair.

In terms of chiral random matrix models, the pertinent ensemble for 
investigating Wilson fermions for QCD is the Gaussian unitary ensemble
for three and more colors and is the Gaussian 
orthogonal ensemble\cite{HALASZJAC} for the case of two colors,
provided that the quarks are in the fundamental representation of color. 
With this in mind, the pertinent random matrix model for spatially constant 
Wilson modes of one-flavor on a cylinder is 
\be
{\bf Q}_W = \left(
          \left( \begin{array}{cc} m_f & \partial_{\tau}\\
                                   -\partial_{\tau} & - 
           m_f \end{array}\right) + 
           \raisebox{-0.5ex}{\mbox{\Large $\displaystyle \bf R$}}
	\right)  \,,
\label{222}
\ee
where $\partial_{\tau}$ is the $\tau$-derivative on a cylinder of length 
$\beta= 1/T$, and ${\bf R}$ a random matrix.
 Equation~(\ref{222}) is defined
with anti-periodic boundary conditions (\ref{matsu}).

The operator (\ref{222}), rewritten in the basis (\ref{matsu})
 is a the sum of a deterministic and a random piece,
\be 
{\bf Q}_W = \left(\matrix{ m &  i{\bf \Omega} \cr
-i{\bf \Omega}  & - m \cr}\right) + 
  \left(\vphantom{\matrix{ m &  i{\bf \Omega} \cr -i{\bf \Omega}  & - m \cr}}
    \, \raisebox{-0.5ex}{\mbox{\Large $\displaystyle \bf R$}}\, \right)
	\,.
\label{xxdirac}
\ee
Due to the presence of Wilson r-terms, the Dirac 
operator is only hermitean, the chiral structure is lost. In contrast
to ${\bf Q}_S$ in (\ref{XDIRAC}), 
${\bf Q}_W$ in (\ref{xxdirac}) is not block-off-diagonal.
The resolvent for the operator (\ref{xxdirac}) is defined as
\be 
G(z)= \frac{1}{2NN_*}\left< {\rm Tr} \frac{1}{z-{\bf Q}_W} \right> \,,
\label{3}
\ee
where $N_*$ is the number of pairs of Matsubara frequencies retained.
The averaging depends on the $\tau$ dependence of the random matrix ${\bf R}$, 
as discussed above in the context of the NJL model. Two cases will be 
considered in this paper : (I) The case where ${\bf R}$ is $\tau$-dependent,
in which case the various Matsubara modes can mix. (II) The case where the 
matrix ${\bf R}$ is $\tau$-independent, with a block diagonal structure
\be
{\bf R}= \otimes {\bf R}_n
\label{blockchiral}
\ee
in which case, the various Matsubara frequencies are decoupled.

Throughout, and for simplicity, we choose Gaussian weights 
$V(a)=\frac{1}{2}a^2$ for the random distributions
\be
\left< \ldots \right> =\frac{1}{Z} \int \ldots \exp[-2N\Sigma^2 {\rm Tr 
	V({\bf R})}]\ d{\bf R} \,,
\label{weight}
\ee
although certainly higher (polynomial) weights are possible \cite{BREZIN1}.
Note that since the dimension of the Gaussian matrix
appears explicitly in the random weight, the widths of the 
Gaussians for the mixed and non-mixed case differ and are related by
the factor $\sqrt{N_*}$. In Sections~4 and 5 we use $\Sigma_Q^2=1$ and
$\Sigma_C^2=1$ for the Quinto and Cardano classes, respectively.
In Section~6, in comparing the two classes we set $2\Sigma_Q^2 = \Sigma_C^2=1$.

The eigenvalue  distribution of the Dirac operator (\ref{xxdirac}) 
is related to the discontinuity of $G(z)$ through the real axis
\be 
\nu(\lambda )=-\frac{1}{\pi}
\,{\rm Im} G(z=\lambda +i 0) \,.
\label{5}
\ee
Using either the law of addition of random matrices \cite{ZEE} or
diagrammatic techniques \cite{BREZINZEE}, it follows that the resolvent $G(z)$
is given by the solution of the algebraic equation (Pastur equation)
\cite{PASTUR}. When the Matsubara frequencies are mixed (model I), the 
corresponding Pastur equation reads
\be
G(z)=\frac{1}{2N_*}\sum_{i=1}^{2N*} \frac{1}{z-G(z) -\epsilon_i}
	\,,
\label{Pastur1}
\ee
where $\epsilon_i$ are the eigenvalues of the deterministic part of the 
Dirac operator (\ref{XDIRAC}). The equation is a polynomial of
$2N_*+1$ order in $G(z)$. When the Matsubara frequencies are decoupled (model 
II), the corresponding Pastur equation is
\be
G(z)=\frac{1}{N_*} \sum_{n=1}^{N_*} G_n(z) \,,
\label{Pastur2}
\ee
where each of the $G_n(z)$ satisfies a  cubic (Cardano class) equation 
\be
G_n(z)= \frac 1{2}
\left(
\frac{1}{z-G_n(z)- {\bf M}_n}+\frac{1}{z-G_n(z)+ {\bf M}_n}\right) \,.
\label{6}
\ee
We note that $M_n (T) =\sqrt{ m^2 + \omega_n^2}$, in overall agreement
with dimensional reduction arguments from high temperature QCD 
\cite{ZAHEDHANS,DETAR}.

\section{Coupled Wilson Fermions}
For Wilson fermions, we consider (\ref{xxdirac}) in the case of two 
lowest Matsubara frequencies. All our considerations, will be made for a 
single 
flavor. For the coupled case (model I), the Dirac operator reads explicitly
\be
{\bf Q} = \left( \begin{array}{cccc} m & 0 & i\omega_0 & 0 \\
			0 & m & 0 & i\omega_1 \\
			-i\omega_0 & 0 & -m & 0 \\
                        0 & -i\omega_1  & 0 & - m \end{array}\right)  +
          \left(\vphantom{\begin{array}{c} 0 \\ 0 \\ 0 \\ 0 \end{array}}\,
	  \raisebox{-1ex}{\mbox{\Huge $\displaystyle R$}}\ \,\right) \,.
\label{q5def}
\ee
The resolvent of the {\em deterministic part} for the two lowest pairs 
of frequencies ($\omega_{-n}=-\omega_{n-1}$) reads
\be
G_D = \alpha \frac{z}{z^2 - M_0^2} + (1-\alpha) \frac{z}{z^2 - M_1^2}
\ee
with
\be
M_n^2 = m^2 + ((2n+1) \pi T)^2
\ee
Here $n=0,1$ and $\alpha$ (hereafter  equal to $1/2$) refers to the 
relative weight of the modes. Adding the
corresponding Blue's functions of the deterministic and the random part one
arrives at a fifth order algebraic equation\cite{MESON96}
\begin{figure}[tbp]
\centerline{\epsfxsize=7.5cm \epsfbox{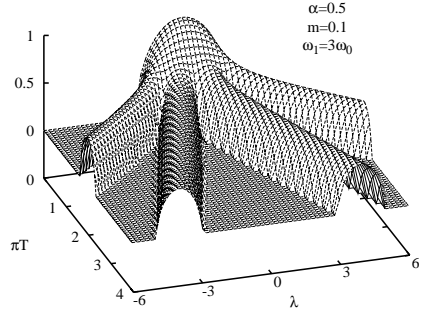}}
\caption{Spectral function for Quinto class for the lowest two Matsubara
frequencies with small mass $m=0.1$ in units of $\Sigma_Q = 1$}
\label{fig1}
\end{figure}

\be
G^5 + a_4 G^4 + a_3 G^3 + a_2 G^2 + a_1 G + a_0 = 0
\label{quintoeq}
\ee
with the following coefficients
\be
a_4 &=& -4 z \nonumber \\
a_3 &=& 1 - (M_0^2 + M_1^2) + 6 z^2 \nonumber  \\
a_2 &=& -4 z^3 -3 z + 2 z (M_0^2 + M_1^2) \\
a_1 &=& \alpha (M_0^2 - M_1^2) - M_0^2 (1-M_1^2)
    + z^2 \left[ 3 - (M_0^2 + M_1^2) \right] + z^4 \nonumber  \\
a_0 &=& z \left[ M_0^2 - \alpha (M_0^2 - M_1^2) - z^2 \right] \,.\nonumber
\ee

\begin{figure}[tbp]
\centerline{\epsfxsize=7.5cm \epsfbox{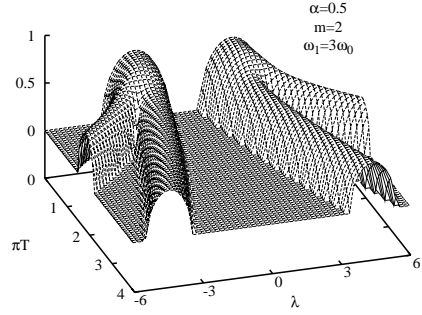}}
\caption{Spectral function for Quinto class for the lowest two Matsubara
frequencies with large mass $m=2$ in units of $\Sigma_Q = 1$}
\label{fig2}
\end{figure}

The spectral function generated by the solutions to (\ref{quintoeq}) through 
(\ref{5}) allow for a rich phase structure (Quinto class). In general, four
possible phases $P_1,P_2,P_3,P_4$ defined by the number of allowed
disconnected arcs or supports of the eigenvalue distribution, are possible.
In Fig.~\ref{fig1} we show the distribution of eigenvalues as a function 
of temperature and for a single light quark flavor of mass $m=0.1$ in units
of $1/\Sigma_Q$.  At zero temperature, the spectral function is peaked 
around zero virtuality, with a nonzero condensate. The system is in the $P_1$
phase, and the distribution is Wigner's semicircle. As the temperature 
increases, the condensate decreases, followed by the decoupling of the 
Matsubara modes. The heaviest, the first. The structural changes are :
$P_1  \rightarrow P_2 \rightarrow P_4$ or $P_1 \rightarrow P_3 \rightarrow P_4$
 or $P_1 \rightarrow P_4$, depending on whether the  restoration
 of chiral symmetry  precedes, follows or parallels the splitting of the 
 frequencies, respectively.  Figure~\ref{fig2} shows the behavior of the 
spectral distribution for a heavier quark flavor. In this case, the quarks are 
always localized with no condensate whatever the temperature. An increase in 
the temperature causes only a decoupling of the two Matsubara modes, that is a 
structural phase change from  $P_2 \rightarrow P_4$.
 
Since the general solutions of a fifth order algebraic equation  do not exist
in analytical form, the results shown in  Figs.~\ref{fig1},\ref{fig2} were 
obtained numerically. Out of the five possible solutions, four are ruled out
by enforcing normalizability and positivity of the spectral distribution.
To check that the presented solution is indeed correct and unique, we have
constructed analytically the equations for the evolution of the endpoints.
These equations are best amenable in the form of Blue's functions,
i.e. the functional inverses of the resolvents: $ G[B(z)] =z$.
Due to the cuts of the Green's functions, the evolution equations
stem from the condition $B'(z)|_{e.p.} =0$, where prime denotes
differentiation with respect to $z$ and the derivative is taken at 
the endpoint (e.p.). Pastur equations rewritten in terms of the 
Blue's functions are always one degree lower than those for
the Green's functions \cite{ZEE}, therefore the evolution equations
for the endpoints in our case are solution to a forth order (Ferrari)
equation.

\begin{figure}[tbp]
\centerline{\epsfxsize=7.5cm \epsfbox{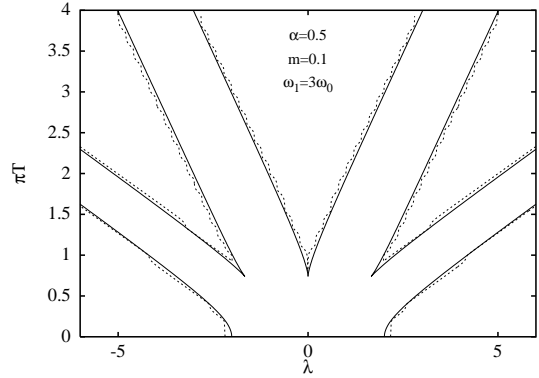}}
\caption{Evolution of the endpoints of the spectral function following 
numerically from (\protect\ref{quintoeq}) (dashed curve), and analytically 
from (\protect\ref{endpoints}) (solid lines)}
\label{fig3}
\end{figure}
In the case considered here,  
the  endpoints $\pm A_i$ $(A_1>A_2>A_3>A_4)$ of the four arcs 
$[-A_1,-A_2]$, $[-A_3,-A_4]$, $[A_4,A_3]$, $[A_2,A_1]$ 
are explicitly given by
\be
A_i= \sqrt{X_i} \frac{X_i^2 - (M_0^2+M_1^2-1)X_i+
\label{endpoints}
  F(\alpha,M_0^2,M_1^2)}{(X_i- M_0^2)(X_i-M_1^2)} \,,\\
F(\alpha,M_0^2,M_1^2) = \alpha(M_0^2-M_1^2)-M_1^2
+M_0^2 M_1^2 \nonumber
\label{endferrari}
\ee
where $X_i$ are real positive roots of the Ferrari equation
\be
X^4-b_3 X^3 -b_2 X^2 -b_1 X- b_0 =0
\label{Ferr}
\ee
 with coefficients
 \be
 b_3&=&1+2(M_0^2+M_1^2) \nonumber \\
 b_2&=&3\alpha(M_0^2-M_1^2)-(M_0^2+M_1^2)^2-2M_0^2(1+M_1^2 )\nonumber \\
 b_1&=&-\alpha (M_0^4-M_1^4)+2M_0^2M_1^2(M_0^2+M_1^2+1)+M_0^4 \nonumber \\
 b_0&=&-\alpha M_0^2 M_1^2(M_0^2-M_1^2) +M_0^4M_1^2(1-M_1^2) \,.
 \label{coeforferr}
 \ee
 The critical points are determined from the condition that the real positive
 points $A_4, A_3, A_2$ melt or vanish. In  Fig.~\ref{fig3} we show the 
 evolution of the endpoints for the numerically generated solutions above 
(dashed curve) versus the analytically generated solutions (solid curve)
from  (\ref{endpoints}). Numerically, the endpoint curve was generated 
using the limit of the norm $|{\rm Im } G(z) -\epsilon |\rightarrow 0$, with 
$\epsilon =0.01$ and $G(z)$ the normalizable solution to (\ref{quintoeq}).
The agreement of the two curves, confirms the uniqness of the spectral 
distribution discussed above.

\section{Decoupled Wilson Fermions}
\begin{figure}[tbp]
\centerline{\epsfxsize=7.5cm \epsfbox{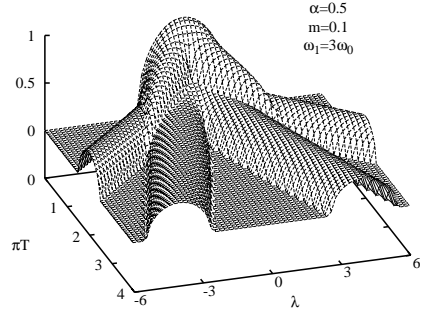}}
\caption{Spectral function for model II and a light mass $m=0.1$}
\label{fig4}
\end{figure}

\begin{figure}[tbp]
\centerline{\epsfxsize=7.5cm \epsfbox{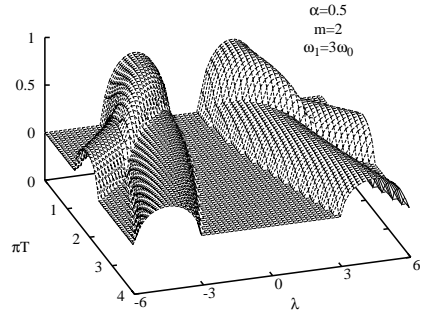}}
\caption{Spectral functions for model II and a heavy mass $m=2$}
\label{fig5}
\end{figure}

The case of Wilson fermions with two lowest but decoupled Matsubara modes 
(model II), follows from (\ref{Pastur2}) through
\be
G(z) &=& \frac{\alpha}{2} G_0(z) + \frac{(1-\alpha)}{2} G_1(z)
\label{matar}
\ee
with each $G_n(z)$ ($n=0,1$) 
satisfying the cubic Pastur equation~\cite{PASTUR}
\be
G_n (z) = \frac{z-G_n(z)}{(z-G_n(z))^2 - M_n^2} \,.
\ee
The evolution equations for the endpoints 
in this case come from the overlap of individual supports for each frequency. 
Since for each frequency the equation for the resolvent is cubic, 
the corresponding equation for the endpoints is quadratic, with the following 
positive solutions
\be
A_n^{1,2}=\frac{1}{\sqrt{2}}
\cdot \frac{3\pm \sqrt{1+8M_n^2}}{1\pm \sqrt{1+8M_n^2}}
\cdot \sqrt{1+2M_n^2\pm\sqrt{1+8M_n^2}}
\label{quadraticendpoint}
\ee
for $\alpha =1/2$. 
Figures~\ref{fig4} and~\ref{fig5} show the spectral functions versus 
temperature, for a light and heavy quark flavor in units of $1/\Sigma_C$.

\section{Comparison of the Models}
Despite the overall structural similarities between the spectral functions 
discussed for model I and II, there are a number of quantitative differences 
upon closer look. These differences may be helpful for distinguishing
these two  
scenarios in a lattice simulation for instance.

\begin{figure}[tbp]
\centerline{\epsfxsize=7.5cm \epsfbox{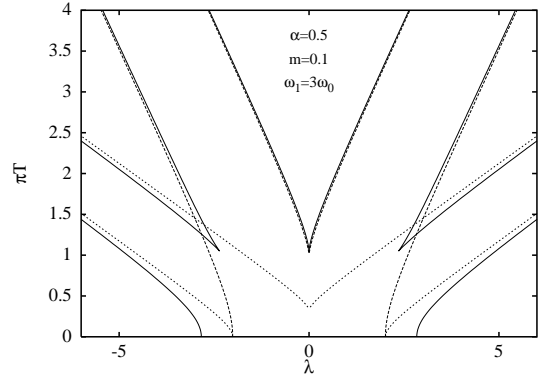}}
\caption{Evolution of the endpoints of the spectral distribution for model I
(solid curve) and model II (dashed curve) with a light quark mass$m=0.1$}
\label{fig6}
\end{figure}

The evolution of the endpoints of the spectra as obtained from model I 
and II are different. Figure~\ref{fig6} shows the behavior of the endpoints 
versus temperature for I (solid line) and II (dashed line) for the light quark 
flavor case $m=0.1$ in units where $2\Sigma_Q^2=\Sigma_C^2=1$. The solid
curves  
follow from (\ref{quadraticendpoint}), while the dashed curves follow from 
(\ref{endpoints}). At high temperature ($T \geq 3 T_c$), the evolution
equation  
becomes degenerate. This is expected following the thermal
localization. Around  
the critical temperature, however, the evolution of the endpoints are 
qualitatively different.

\begin{figure}[tbp]
\begin{minipage}[t]{6.5cm}
{\epsfxsize=6cm \epsfbox{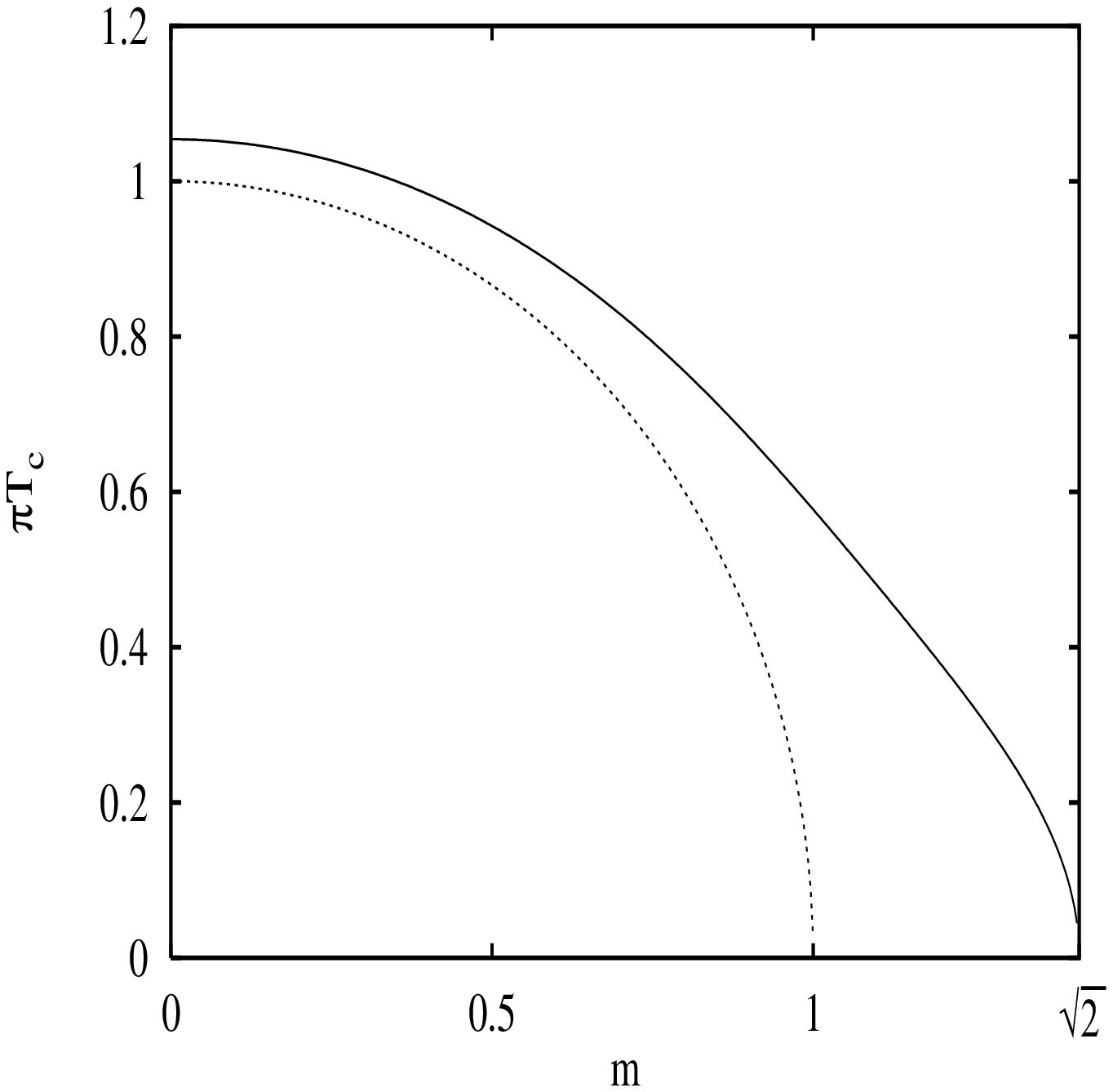}}
\caption{Critical temperature as a function of the current mass $m$
for model I (solid curve) and model II (dashed curve).
The units are fixed by $\Sigma_C^2=2\Sigma_Q^2=1$}
\label{fig5a}
\end{minipage} \hfill
\vspace*{-8.7cm} \hspace*{6.5cm} \hfill
\begin{minipage}[t]{6.5cm}
{\epsfxsize=6cm \epsfbox{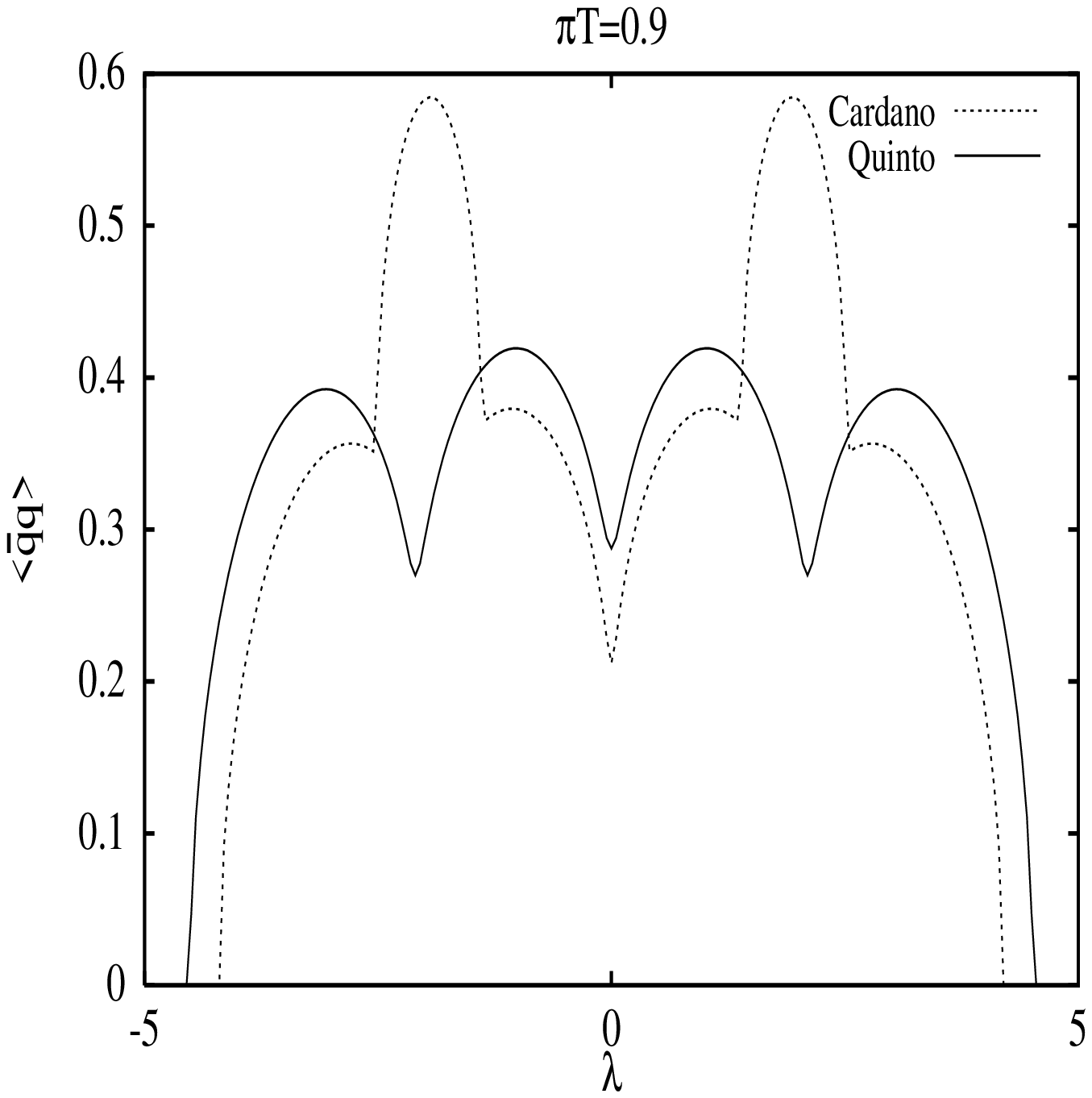}}
\caption{Comparison of the spectral functions for model I(solid line),
and model II (dashed line) with
$m=0.1$ at $\pi T=0.9$. The units are fixed by $\Sigma_C^2 =
2\Sigma_Q^2$ for comparison}
\label{fig7}
\end{minipage}
\end{figure}

The critical masses and temperatures, are of course different for the two 
cases considered. For model I and at zero virtuality
($\lambda = 0$) the Quinto equation reduces to a 
second order equation, with the solution
\be
G^2 = \frac{\Sigma_Q^2 (M_0^2+M_1^2)-1}{2} \pm \frac{1}{2} 
  \sqrt{1+\Sigma_Q^4 (M_0^2-M_1^2)^2} 
\ee
for $\alpha =1/2$. 
The support for the spectral density vanishes when $G^2$ becomes real and
positive, that is for
\be
M_0^2+M_1^2-2\Sigma_Q^2 M_0^2 M_1^2 = 0 .
\ee
This yields the critical temperature
\be
&4 \pi^2\Sigma_Q^2 T_c^2 = \frac{10}{9} (1-2\Sigma_Q^2 m^2) + \nonumber \\
&  \sqrt{ \frac{100}{81} (1-2\Sigma_Q^2 m^2)^2 -
  \frac{16}{9} \Sigma_Q^2 m^2 (\Sigma_Q^2 m^2 -1)} .
\label{tcrit}
\ee
For zero quark mass (\ref{tcrit}) gives $\pi^2 T_c^2  = {5}/{9\,\Sigma_Q^2}$.
The critical mass, at which the condensate vanishes at zero temperature
is $m_c = {1}/{\Sigma_Q}$. Figure~\ref{fig5a} shows the behavior of the
critical  
temperature versus $m$ for model I or Quinto class (solid curve). The dashed 
curve is the expected result from model II, with
\be
\pi^2T_c^2 = 1-m^2 \,.
\ee
For the massless case the critical temperature is in this case $\frac{1}{\pi}$,
while at zero temperature the critical mass is 1.  The comparison between the 
two models, was carried for $2\Sigma_Q^2=\Sigma_C^2=1$. Note that the ratio
of the critical parameters depend on the mixing scenario
\be
\frac{\pi T_c}{m_c} = \left\{ \begin{array}{ll}
  1 & \mbox{\,\,\,model I} \\
  \sqrt{\frac{5}{9}} & \mbox{\,\,\,model II} \,.
\end{array} \right.
\ee

The spectral distributions around the critical temperature are also 
qualitatively different for the two models considered, as shown in 
Fig.~\ref{fig7}. The spectral function in model I following from the 
Quinto equation (solid curve) is smooth throughout, while the one for
model II following from the superposition of two Cardano solutions
(dashed curve) has sharp edges (discontinuous derivatives). The wiggly 
behavior seen in both spectra is reminiscent of a similar feature seen in 
Kalkreuter lattice spectra with Wilson fermions \cite{KALKREUTER}. 
This is suggestive of some additional (finite size) scale in the latter.
This point is worth investigating.

Finally, in both models the critical exponents are mean-field 
\cite{JACKSON,USNJL,BROWNRHOBUB}. This is expected 
in large $N$ and for Gaussian randomness for all polynomial classes
(Cardano, Ferrari, Quinto and higher)\footnote{Quadratic equations
do not allow for a phase transition. The system is always in the $P_1$ phase
(Wigner semicircle).}

\section{Staggered Quark Spectra}
For chiral fermions (\ref{XDIRAC}) with decoupled Matsubara frequencies, 
consider the case with the dimensional reduction argument~\cite{ZAHEDHANS}
\be
{\bf Q}_S^{II} = \otimes_{\pm,n}\ {\bf Q}_{S,n}^{II}
= \otimes_{\pm,n}\ 
  \left( \begin{array}{cc} 0 & \pm M_n(T) \cr
  \pm M_n(T) & 0 \end{array}\right) \ + \ 
  \left( \begin{array}{cc} 0 & R \cr
  R^{\dagger} & 0 \end{array}\right)
\label{omega2}
\ee
with $M_n(T) = \sqrt{m_s^2+\omega_n^2}$.
The spectral distribution associated to (\ref{omega2}) follows from the 
combination of discontinuities of the pertinent solution to a 
cubic (Cardano) equation, Eqs.~(\ref{Pastur2},\ref{6}). The case
considered here at  
high enough temperature is equivalent to the one of Wilson fermions with
decoupled Matsubara modes.

Following Chandrasekharan and Christ \cite{CHRIST}, let us define the 
valence quark condensate,
\be
<\overline{\zeta}\zeta > = 2m_{\zeta} \int_0^{\infty} d\lambda 
\frac{\nu (\lambda)}{\lambda^2 + m_{\zeta}^2} \,,
\label{CC1}
\ee
where $m_{\zeta}$ is the valence quark mass, and $\nu (\lambda)$ the spectral
density associated to (\ref{omega2}) for massless sea quarks, that is 
$m_s=0$. Although we have discussed a single quark flavor, the analysis 
carries through unchanged in the massless case. 
\begin{figure}[tbp]
\centerline{\epsfxsize=6cm \epsfbox{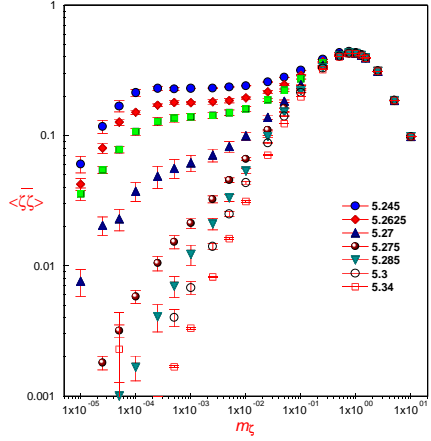}}
\caption{Semi-quenched condensate versus the valence quark mass $m_{\zeta}$
for two flavor QCD \protect\cite{CHRIST} with a sea mass $m_sa =0.01$}
\label{fchrist}
\end{figure}

\begin{figure}[tbp]
\begin{minipage}[t]{6.5cm}
{\epsfxsize=6cm \epsfbox{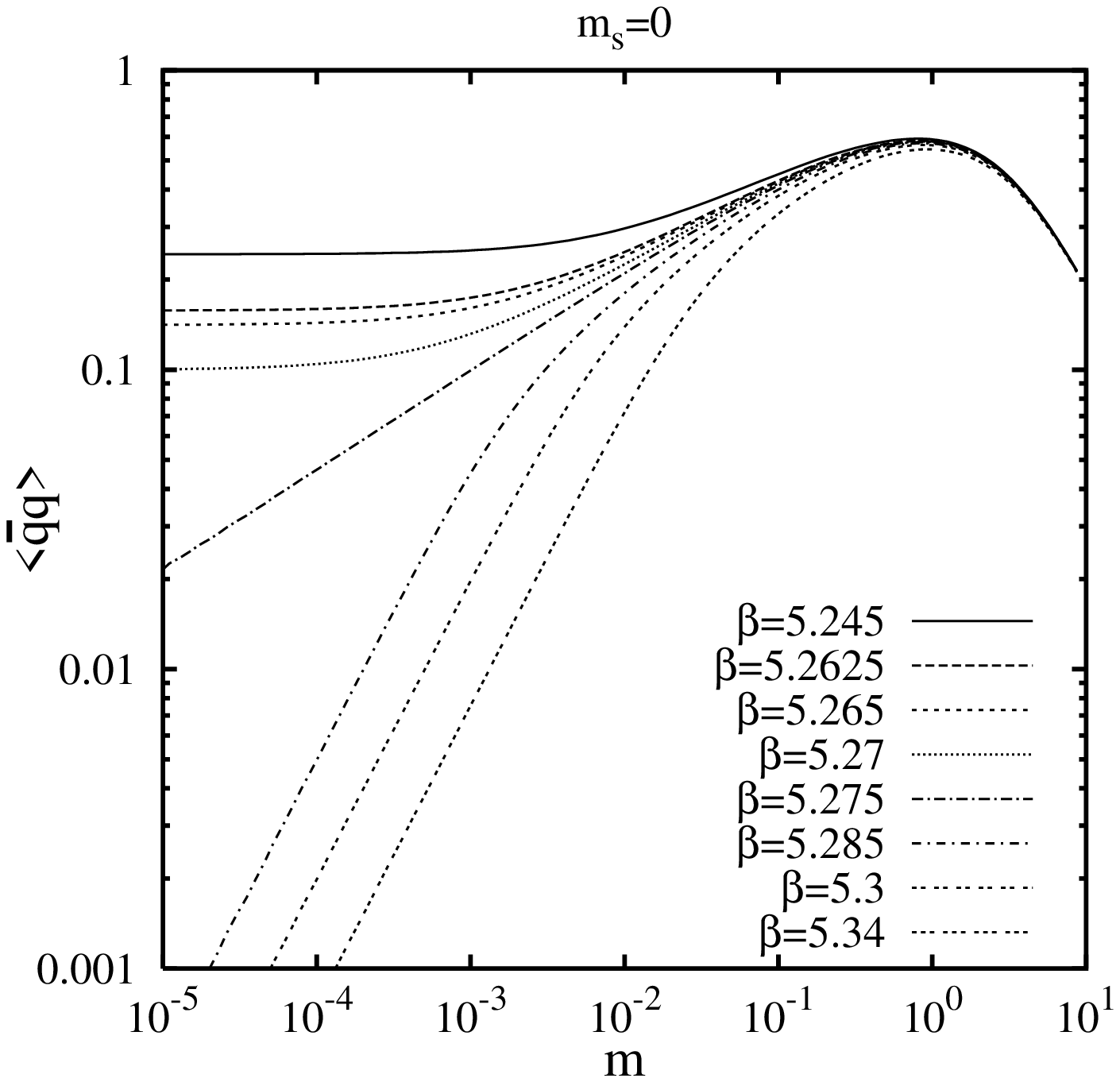}}
\end{minipage} \hfill
\begin{minipage}[t]{6.5cm}
\centerline{\epsfxsize=6cm \epsfbox{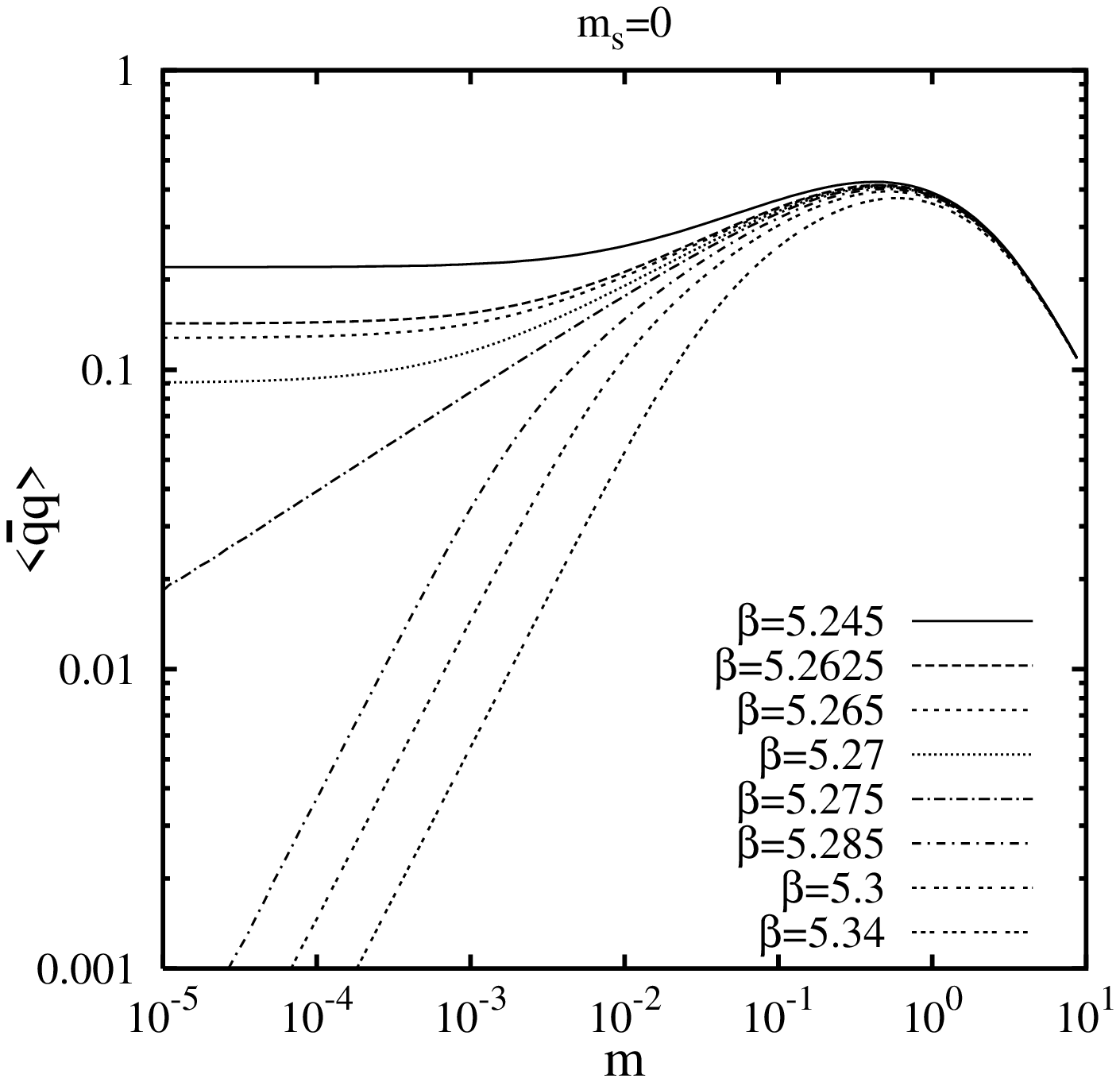}}
\end{minipage}
\caption{Semi-quenched condensate versus the valence quark mass $m_{\zeta}$
obtained from model II (\protect\ref{CC1}) for the lowest Matsubara modes 
(left) and the one corresponding to the coupling (model I) (right) with
zero sea mass}
\label{fch-car1}
\end{figure}

In Fig.~\ref{fchrist} we show the behavior of the valence quark condensate 
as obtained by Chandrasekharan and Christ \cite{CHRIST}, using 
two-flavor QCD with staggered 
fermions and $m_s a =0.01$ (or $m_s=6$ MeV in physical units). 
In Fig.~\ref{fch-car1} we
display the results following from (\ref{CC1}) with $m_s=0$, with the two
lowest Matsubara pairs (right figure) and for comparison
the same result but for one Matsubara mode (left figure)
as discussed in \cite{WILSON}. 
The results are in 
overall agreement with each other, showing mild sensitivity to $m_s$ and 
the addition of further Matsubara modes. The largest sensitivity appears 
around large values of $m_{\zeta}$ in units of $1/\Sigma$. 

Model I for the mixing of the two lowest Matsubara modes is not trivial
in the case of chiral fermions. Since for the decoupled case model II
was equivalent to the one of Wilson fermions, we used as model I for chiral
fermions the spectral density obtained from (\ref{q5def}) with the
proper scaling $2\Sigma_Q^2 = \Sigma_C^2$.
Comparing the large mass part of the spectra we may conclude that the
model with the coupling between the Matsubara modes agrees better with 
the lattice calculation (Fig.~\ref{fchrist}).

\section{Conclusion}

We have investigated the role of few Matsubara modes on random matrix models
as inspired from QCD spin and flavor symmetry. After showing how the mixing 
between the various Matsubara modes may affect, the generic structure of the 
random matrix formulation at finite temperature, we have considered two 
complementary models. One with mixing (I) and the other without (II). The
spectral function associated to the former was shown to follow from a quintic
equation (Quinto class), while the one for the latter was shown to follow from 
the superposition of two solutions to quartic equations (Cardano class). 

For a light quark mass, the spectral distribution at low temperature has a 
support at zero virtuality. Randomness prevails over temperature. As the 
temperature is increased, the thermal effects cause the spectral distribution 
to vanish. Thermal localizations are found to follow the thermal wavelengths. 
For a heavy quark mass, the spectral distribution shows no support at zero 
virtuality even at zero temperature owing to the localization over the Compton 
wavelength. With increasing temperature, the thermal localization takes over,
with a segregation along the thermal frequencies.

While schematic, the present analysis shows that around the critical 
temperature the effects of the higher Matsubara modes cause quantitative 
changes not only in the critical parameters and their ratios, but also in the
endpoints evolution and the spectral shapes. These observations are of 
relevance for lattice simulations of Dirac spectra for both Kogut-Susskind 
and Wilson fermions. 


\vglue 0.6cm
{\bf \noindent  Acknowledgments \hfil}
\vglue 0.4cm
I thank Maciej A. Nowak and Ismail Zahed for the valuable discussions.
This work was supported by 
the Hungarian Research Foundation OTKA grants T022931 and F019689.

\vglue 0.6cm
{\bf \noindent  Notes \hfil}
\vglue 0.4cm
\begin{flushright}
{\it Dedicated to G. Marx and K. Nagy on the occasion of their 70th
	birthdays}
\end{flushright}

\vskip 1cm
\setlength{\baselineskip}{15pt}

\end{document}